# Heat Transfer Simulation for Optimization and Treatment Planning of Magnetic Hyperthermia Using Magnetic Particle Imaging

Natsuo Banura, Atsushi Mimura, Kohei Nishimoto, Kenya Murase*

Department of Medical physics and Engineering, Division of Medical Technology and Science,
Course of Health Science, Graduate School of Medicine, Osaka University, Suita, Osaka 565-0871, Japan
*Corresponding author, email: murase@sahs.med.osaka-u.ac.jp

**Abstract**
This study was undertaken to develop a system for heat transfer simulation for optimization and treatment planning of magnetic hyperthermia treatment (MHT) using magnetic particle imaging (MPI). First, we performed phantom experiments to obtain the regression equation between the MPI pixel value and the specific absorption rate (SAR) of magnetic nanoparticles (MNPs), from which the MPI pixel value was converted to the SAR value in the simulation. Second, we generated the geometries for use in the simulation by processing X-ray computed tomography (CT) and MPI images of tumor-bearing mice injected intratumorally with MNPs (Resovist®). The geometries and MPI images were then imported into software based on a finite element method (COMSOL Multiphysics®) to compute the time-dependent temperature distribution for 20 min after the start of MHT. There was an excellent correlation between the MPI pixel value and the SAR value (r = 0.956). There was good agreement between the time course of the temperature rise in the tumors obtained by simulation and that obtained experimentally. The time-dependent temperature distributions under various blood perfusion rates in the tumors and at various locations could be simulated, which cannot be easily obtained experimentally. These results suggest that our system for heat transfer simulation using MPI will be useful for the optimization and treatment planning of MHT.

## I. Introduction

Recently, a new imaging method called magnetic particle imaging (MPI) has been introduced [1]. MPI allows imaging of the distribution of magnetic nanoparticles (MNPs) with high sensitivity, high spatial resolution, and high imaging speed. In addition, MPI can visualize MNPs in positive contrast with no signals from background tissues and can quantify the amount of MNPs with excellent linearity. More recently, we have developed a system for MPI with a field-free line (FFL) encoding scheme using two opposing neodymium magnets and a gradiometer-type receiving coil [2].

Magnetic hyperthermia treatment (MHT) is a strategy for cancer treatment using the temperature rise of MNPs as a result of their hysteresis and relaxational losses under an alternating magnetic field (AMF) [3]. This strategy aims to heat only tumors by accumulating the MNPs within them using a drug delivery system. Accurate knowledge of the spatial distribution and amount of MNPs accumulated in the targeted region is crucial for designing optimal treatment planning of MHT to prevent insufficient heating of the targeted region and overheating of the healthy tissue.

The purpose of this study was to develop a system for heat transfer simulation for optimization and treatment planning of MHT using MPI.

## II. Materials and Methods

### II.I. Phantom experiments

We performed phantom experiments to investigate the relationship between the MPI pixel value and the specific absorption rate (SAR) of MNPs [4]. In this study, we used Resovist® as the source of MNPs. First, we prepared samples by putting Resovist® into a cylindrical polyethylene tube 6 mm in diameter and 5 mm in length (100 μL) with various iron concentrations of Resovist® (0, 50, 100, 125, 250, and 500 mM) and imaged them using our MPI scanner [2]. Transverse images were reconstructed from the projection data using the maximum likelihood-expectation maximization (ML-EM) algorithm over 15 iterations. We defined an average MPI value as the mean of the pixel values within a circular region of interest (ROI) with a diameter of 6 mm, drawn on the transverse images. These samples were also heated using our apparatus for MHT [5] by applying the AMF at 600 kHz and 3.1 kA/m for 10 min, and the time course of the temperature rise was measured using an infrared thermometer (FLIR E4, FLIR Systems Inc., OR, USA). In this study, we calculated the SAR value using the following equation:

$$SAR = \rho c \left(\frac{\Delta T}{\Delta t}\right)_0, \quad (1)$$

where $\rho$ is the density of ferrofluid (1290 kg/m³), $c$ is the specific heat of ferrofluid (4100 J/(kgK)), and $(\Delta T/\Delta t)_0$ is the initial slope of the time-dependent temperature rise. $(\Delta T/\Delta t)_0$ was obtained by fitting the time-dependent temperature rise ($\Delta T(t)$) by use of the phenomenological Box-Lucas equation [6] given by

$$\Delta T(t) = T(t) - T(0) = A(1 - e^{-\beta t}), \quad (2)$$

where $T(t)$ and $T(0)$ are the temperatures at time $t$ and 0, respectively, and $A$ and $B$ are constants. Finally, the linear regression equation between the average MPI value and the SAR value was obtained.

## II.II Animal experiments

The details of animal experiments are described in our previous paper [4]. In brief, seven-week-old male BALB/c mice were purchased from Charles River Laboratories Japan, Inc. (Yokohama, Japan). After one-week habituation, Colon-26 cells ($1 \times 10^6$ cells) were implanted subcutaneously into the back of a mouse under anesthesia. When the tumor volume reached approximately 100 mm³, the tumors in the mice were injected with Resovist® at an iron concentration of 250 mM (half-dose) or 500 mM (full-dose). Ten minutes after the injection of Resovist®, MPI images were obtained in the same manner as in the phantom studies [4]. After the MPI studies, X-ray CT images were obtained using a 4-row multi-slice CT scanner with a tube voltage of 120 kV and a tube current of 210 mA. Thirty minutes after the injection of Resovist®, MHT was performed by applying the AMF at 600 kHz and 3.1 kA/m for 20 min. During MHT, the temperatures in the tumor and rectum were measured using two optical fiber thermometers. All of the animal experiments described above were approved by the animal ethics committee at Osaka University School of Medicine.

## II.III. Heat transfer simulation

First, the X-ray CT images obtained in animal experiments were processed using MATLAB® (Natick, MA, USA) codes to generate the geometries of tumor-bearing mice. Four regions (soft tissue, bone, spine, and sinus) were extracted from the X-ray CT images using K-means clustering, and the tumor region was segmented by manually setting an ROI. After these segmentations, the contours of the regions were extracted by an edge detection method, and the edge images were converted into drawing exchange format (DXF) and were imported to a software program based on a finite element method (COMSOL Multiphysics®, Stockholm, Sweden) as the geometry. We also generated ROI images from the MPI images by taking the threshold value for extracting the contour as 40% of the maximum MPI value. After the ROI images were co-registered to the tumor regions, they were imported to COMSOL Multiphysics®. The pixel values in the ROI image were converted to the SAR values using the regression equation obtained by phantom experiments. Finally, Pennes' bioheat equation [7] was solved to simulate the heat transfer in the geometry of a tumor-bearing mouse, which is given by

Tumor
$$\rho_t c_t \frac{\partial T_t}{\partial t} = k_t \nabla^2 T_t + Q_t + SAR + \rho_b c_b \omega_t (T_b - T), \quad (3)$$

Healthy tissue
$$\rho_h c_h \frac{\partial T_h}{\partial t} = k_h \nabla^2 T_h + Q_h + \rho_b c_b \omega_h (T_b - T), \quad (4)$$

where $k$ is the thermal conductivity, $\rho$ is the density, $c$ is the specific heat, $Q$ is the metabolic heat generation rate, $T$ is the temperature, and $\omega$ is the blood perfusion rate. The subscripts $t, h,$ and $b$ represent the parameters in the tumor, healthy tissue, and blood, respectively. The thermophysical properties for the tumor and healthy tissue are summarized in Table 1 [8]. The initial temperatures of the tumor and healthy tissue were set at 36 °C, while the blood temperature was set at 37 °C. The tumor and healthy tissue surfaces were subject to a natural convection boundary condition with an external temperature of 28 °C and a heat transfer coefficient of 3.7 W/(m²K) [8].

**Table 1**: Thermophysical properties used in the simulation

|  | Tumor | Healthy tissue | Blood |
|---|---|---|---|
| Density $\rho$ (kg/m³) | 1045 | 1045 | 1060 |
| Heat capacity $c$ (J/(kg K)) | 3760 | 3760 | 3770 |
| Thermal conductivity $k$ (W/(m K)) | 0.51 | 0.51 | |
| Blood perfusion rate $\omega$ (s⁻¹) | 0.0095 | 0.003 | |
| Metabolic heat generation rate $Q$ (W/m³) | 31872.5 | 6374.5 | |

## III. Results

Figure 1 shows the relationship between the average MPI value (x) and the SAR value (y, W/m$^3$) obtained by phantom experiments. There was an excellent correlation between them (r = 0.956) with a linear regression equation of y = 5.42×10$^4$x + 5.27×10$^4$, from which the MPI pixel value of the tumor was converted to the SAR value in Eq. (3).

The results of animal experiments and simulations for MHT are shown in Figs. 2-4. Figure 2(a) shows a typical example of a fusion image created from the MPI and X-ray CT images of a tumor-bearing mouse injected with Resovist® at an iron concentration of 250 mM, while Fig. 2(b) shows the temperature distribution 20 min after the start of MHT, which was obtained by COMSOL Multiphysics®. Figure 2(c) shows a comparison between the time course of the temperature rise in the tumor after the start of MHT obtained by simulation (○) and that obtained experimentally (●). Figure 2(d) shows the simulation results of the time course of the temperature rise in the tumor for various values of the blood perfusion rate in the tumor.

Figure 3(a) illustrates the locations at which the time course of the temperature rise was calculated. The symbols a, b, c, d, and e show the center of the tumor, the tumor side adjacent to the boundary between the tumor and healthy tissue, the boundary between the tumor and healthy tissue, the healthy tissue side adjacent to the boundary, and the center of the body, respectively. Figure 3(b) shows the time courses of the temperature rise at five different locations shown in Fig. 3(a).

Figure 4(a) shows the simulation results of the temperature distribution 20 min after the start of MHT in the tumor-bearing mouse injected with Resovist® at an iron concentration of 500 mM. Figure 4(b) shows a comparison between the time course of the temperature rise in the tumor injected with Resovist® at an iron concentration of 500 mM (●) and that injected with 250 mM (○).

## IV. Discussion

In this study, we developed a system for heat transfer simulation for optimization and treatment planning of MHT using MPI. To our knowledge, this is the first report to investigate the applicability of MPI to treatment planning of MHT. Our system will provide a convenient tool for the above purposes, because it enables us to investigate a large number of study conditions, which cannot be easily realized by experiments.

Our results of phantom experiments demonstrated that the average MPI value has an excellent correlation with the SAR value (Fig. 1), indicating that MPI can be used for predicting the temperature rise in MHT. As shown in Fig. 1, however, a large scatter of the SAR value was observed. This large scatter appears to be mainly due to the fact that the accuracy of temperature measurement using an infrared thermometer is affected by several factors such as the distance between the camera and the target surface and the angle of camera axis with respect to the target surface norm [4].

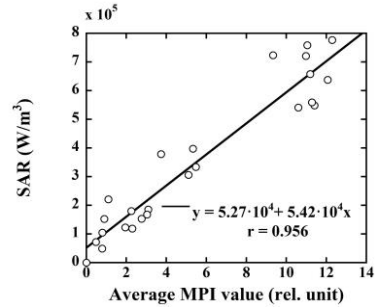

**Figure 1:** Relationship between the average MPI value and the specific absorption rate (SAR).

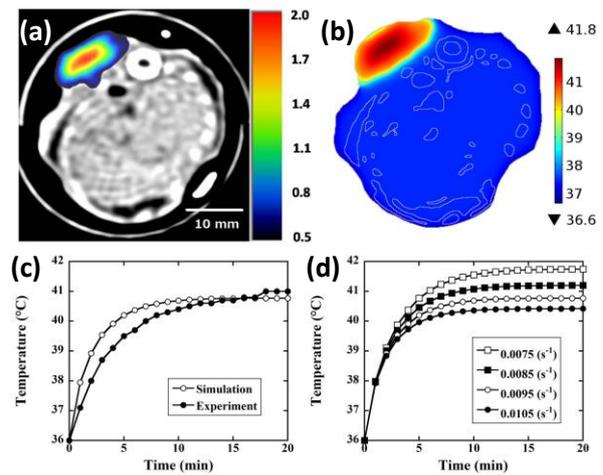

**Figure 2:** (a) Fusion image combining MPI and X-ray CT images of the tumor-bearing mouse injected with Resovist® at an iron concentration of 250 mM. (b) Temperature distribution at 20 min after the start of MHT obtained by simulation, in which the blood perfusion rate $\omega_t$ was fixed at 0.0095s$^{-1}$. (c) Comparison between the time course of the temperature rise in the tumor obtained by simulation (○) and that obtained experimentally (●). (d) Time courses of the temperature rise in the tumor for various values of the blood perfusion rate in the tumor.

As shown in Figs. 2(a) and 2(b), the MPI image of the tumor was very similar to the image of the temperature distribution in the tumor obtained by simulation. In addition, our simulation results (Fig. 4) clearly demonstrated the difference in the time course of the temperature rise between cases when Resovist® was used with iron concentrations of 250 mM (○) and 500 mM (●). These results suggest that the excellent performance of MPI allowing quantitative imaging of MNPs is useful for heat transfer simulation in MHT.

As shown in Fig. 3, the temperature rise within the tumor (a and b) was higher than that outside the tumor

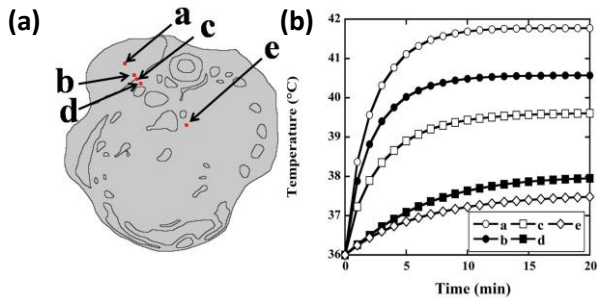

**Figure 3:** (a) Five different locations for simulating the temperature rise (a: center of the tumor, b: tumor side adjacent to the boundary between the tumor and healthy tissue, c: boundary between the tumor and healthy tissue, d: healthy tissue side adjacent to the boundary, and e: center of the body). (b) Time courses of the temperature rise at the locations illustrated in (a).

(d and e), demonstrating that the time course of the temperature rise differs greatly depending on the location, especially the distance from the center of the targeted region in which MNPs exist. Such a regional evaluation of the temperature rise using our system will be useful for preventing insufficient heating in the targeted region and overheating in healthy tissue, resulting in enhancement of the therapeutic effect and the reduction of side effects in MHT.

A limitation of this study is that we used the regression equation obtained by phantom experiments for converting the MPI pixel value to the SAR value. The thermophysical properties of Resovist® when filled in phantoms may differ from those when injected into tumors. It has been reported that the MPI signal changes depending on the factors in MNPs' environment [9] such as viscosity, pH, and aggregation of MNPs. Thus, further studies on the behavior of the MPI signal in different environments and the development of a method for correcting these signal changes will be necessary. Another limitation is that the thermophysical properties of the tumor and healthy tissue used in this study (Table 1) [8] may be different from real ones, because they change depending on the tissue type, temperature, and location [8]. In particular, the blood perfusion rate varies depending on the temperature and extent of the tissue damage caused by heating and therefore the temperature in the tissue will change. Our simulation results (Fig. 2(d)) also demonstrated that the time course of the temperature rise in the tumor varied depending on the blood perfusion rate in the tumor. Thus, it will be necessary to solve Eqs. (3) and (4) by coupling these changes for more detailed simulation. These studies are currently in progress. Furthermore, the development of a reliable and noninvasive method for measuring the thermophysical properties will also be necessary to assure consistent quality of the treatment planning of MHT.

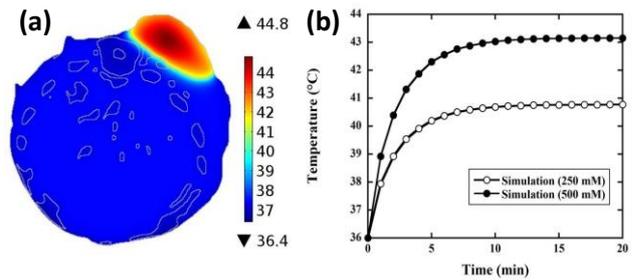

**Figure 4:** (a) Temperature distribution at 20 min after the start of MHT in the tumor-bearing mouse injected with Resovist® at an iron concentration of 500 mM. The blood perfusion rate of the tumor ($\omega_t$) was fixed at 0.0095 s$^{-1}$. (b) Comparison between the time course of the temperature rise in the tumor injected with Resovist® at an iron concentration of 250 mM (○) and that injected with 500 mM (●).

In conclusion, we developed a system for heat transfer simulation in MHT using MPI. Our results suggest that our system using MPI will be useful for the optimization and treatment planning of MHT.

**ACKNOWLEDGEMENT**

This work was supported by a Grant-in-Aid for Scientific Research from the Japan Society for the Promotion of Science (JSPS) and the Japan Science and Technology Agency (JST).